\documentclass[letter,twocolumn]{jpsj3} 

\usepackage{mathrsfs} 
\usepackage{amsmath,amssymb}
\def\CTM{Baxter2} 

\def\newb#1#2{b_{#1,#2}}
\def\newc#1{{c_{#1}}}

\def\AS{\bar{V}}
\def\Aa{{\bf A}} 
\def\Ba{{\bf B}}
\def\Ca{{\bf C}}
\def\Da{{\bf D}}

\def\Cf{{\cal C}}

\def\Df{{\cal D}}
\def\Ff{{\cal F}}
\def\Kf{{\cal K}}
\def\Rf{{\cal R}}
\def\Qf{{\cal Q}}
\def\M1{I}

\def\calL{{\cal L}}
\def\calT{{\cal T}}
\def\Deltax{{a}}

\newcommand{\ket}[1]{\left| #1 \right\rangle}

\newcommand{\bra}[1]{\left\langle #1 \right|}

\def\Tr{{\rm Tr\,}}
\def\bege{\begin{equation}}
\def\ende{\end{equation}}

\def\d#1{\,{\rm d}#1}
\def\e#1{\,{\rm e}#1}

\def\tr#1{\,\mbox{tr }#1}
\def\Tr{\mathop{{\rm Tr}}}

\def\and{\,\,\&\,}

\def\ket#1{|#1\rangle }
\def\bra#1{\langle #1|}

\def\i{{\bf i}}

\def\skipm#1{}
\def\refeq#1{eq.~(\ref{#1})}

\def\i{i}
\title{Continuous Matrix Product Ansatz \\
for the One-Dimensional Bose Gas with Point Interaction
}
\author{Isao \textsc{Maruyama}\thanks{E-mail address: maru@mp.es.osaka-u.ac.jp} and Hosho \textsc{Katsura}$^{1}$\thanks{E-mail address: katsura@kitp.ucsb.edu}}

\inst{Graduate School of Engineering Science, Osaka University, 
Toyonaka, Osaka 560-8531, Japan \\
$^{1}$Kavli Institute for Theoretical Physics, University of California,
Santa Barbara, CA 93106, USA
}
\abst{
We study a matrix product representation of the Bethe ansatz state 
for the Lieb-Linger model 
describing the one-dimensional Bose gas with delta-function interaction. 
We first construct eigenstates of the discretized model in the form of matrix product states using the algebraic Bethe ansatz. 
Continuous matrix product states are then exactly obtained 
in the continuum limit with a finite number of particles.  
The factorizing $F$-matrices in the lattice model are indispensable for the continuous matrix product states 
and lead to a marked reduction from the original bosonic system with infinite degrees of freedom to the five-vertex model.
}
\kword{algebraic Bethe ansatz, matrix product ansatz, continuous matrix product state}

\begin{document}
\maketitle

Finding
an optimal representation of many-body quantum states is the
key issue for variational calculations.
In the density matrix renormalization
group (DMRG) method~\cite{PRL.69.2863},
which is a powerful numerical method for 
one-dimensional 
strongly correlated systems,
the matrix product state (MPS) 
is used as a variational state~\cite{PRL.75.3537}.
In some cases, such as the Affleck-Kennedy-Lieb-Tasaki (AKLT) model~\cite{PRL.59.799},
it is possible to obtain the exact ground state as the MPS.
While the MPS has a long history,~\cite{JSP.9.650,PRB.33.659}
it is currently attracting
interest in the interdisciplinary field of condensed matter physics
and quantum information science~\cite{QIC.7.401}.

At first glance, it seems useless to find 
an optimal representation for eigenstates of quantum integrable models 
since they are exactly obtained from the Bethe ansatz wave functions. 
However, 
direct calculation using the exact eigenstates is a formidable task.
A recent development in this field is the application of the factorizing $F$-matrix (Drinfel'd twist), 
\cite{q-alg.9612012}
which is a similarity transformation to a new basis ($F$-basis) in which actions of operators are simple. 
This new scheme 
enables a direct computation of form factors.~\cite{math-ph.9807020}
The Bethe ansatz has a long and rich history beginning with Bethe's solution of the spin-$\frac{1}{2}$ Heisenberg chain\cite{ZP.71.205}, 
which is now called the coordinate Bethe ansatz, of which there are several variants such as the algebraic Bethe ansatz (or quantum inverse scattering method)\cite{Korepin_textbook} and matrix product ansatz.
The matrix product ansatz proposed by Alcaraz and Lazo
\cite{Alcaraz}
gives exact eigenstates expressed as the MPS. 
Moreover, they have found algebraic relations between matrices constituting MPS,
which reproduce the correct results obtained by the coordinate Bethe ansatz. 
That is, permutations of the
spins in the coordinate Bethe ansatz are expressed by commutation
relations of the matrices.

In this sense, an MPS-type representation of the Bethe states
is interesting in many fields of physics ranging from mathematical physics to quantum information theory. 
Motivated by this, we have studied the connection between the algebraic Bethe ansatz and the matrix product ansatz
and have shown their equivalence in the Heisenberg chain\cite{AX.0911.4215}.
Surprisingly, the connection turns out to be related to the $F$-matrices.
We have found that an MPS-type representation of the Bethe state itself 
is easily obtained by changing the order of the product of 
the $L$-operators.
The concept is similar to 
the quantum transfer matrix
\cite{PTP.56.1454}
and will be explained in detail in the present letter.
To find algebraic relations among matrices appearing in MPS, 
the obtained MPS-type representation is too complicated. 
Thus, the $F$-matrices are necessarily required. 
In the new basis introduced by the $F$-matrices, the matrices 
have a simple structure~\cite{AX.0911.4215}. This simplification corresponds to a mapping 
from the six-vertex model to a five-vertex model.

Recently, Verstraete and Cirac have proposed the use of continuous matrix product states (cMPS)~\cite{AX.1002.1824}
as variational states for one-dimensional continuum models including the Lieb-Liniger model\cite{PR.130.1605}. 
It is a natural extension of the lattice MPS. The Lieb-Liniger (or the quantum nonlinear
Schr\"odinger) model is a model of a nonrelativistic quantum field
theory and describes the one-dimensional Bose gas with the two-body interaction
$V(x,y)=\kappa \delta(x-y)$.
Theoretical studies of the Lieb-Liniger model,
such as the exact calculation of correlation functions\cite{PRL.98.150403},
have received considerable attention
owing to the recent experimental realization of the trapped one-dimensional Bose gas\cite{NATURE.429.277,SCIENCE.305.1125}.
This model also has a long history as one of the exactly solvable models and 
has several different lattice regularizations.  
While all of these models describe the Bose gas at low densities, some of them are non-integrable such as the
Bose-Hubbard model obtained by a natural discretization.~\cite{cond-mat.0405155}
Integrable lattice discretizations can be constructed from the $R$-matrix,
which ensures the integrability through the Yang-Baxter relation.

In this letter, following our previous paper~\cite{AX.0911.4215} , we 
construct a cMPS of the Bethe state 
for the Lieb-Liniger model by taking the continuum limit of the MPS of the Bethe state in the integrable lattice model. 
It is known that 
this quantum lattice Hamiltonian obtained by the discretization 
becomes quasi-local, and includes long-range interactions
that disappear in the continuum limit~\cite{NPB.205.401,TMP.57.1059}.  
One important consequence of this letter is that 
the $F$-matrices are indispensable for explicit representation of cMPS,
while lattice MPS can be obtained without them.
Simplification due to  the $F$-matrices
leads to the five-vertex model for the Lieb-Liniger model,
as for the Heisenberg chain.
To clarify the physical meaning 
in the explicit form of the cMPS,
we introduce a ``world line'' representation of cMPS
as an analog of
the world line in the continuous time loop algorithm\cite{PRL.77.5130}.
We also comment on the cMPS defined by Verstraete and
Cirac\cite{AX.1002.1824}.  In this comparison, we show that the
boundary operator\cite{PRL.75.3537}  plays an important role in fixing the
number of particles in the Bethe state.

Let us start from the Hamiltonian of the Lieb-Liniger model, which is given by 
\begin{eqnarray*}
H= \int_{0}^L  [\partial_x \psi^\dagger(x) \partial_x \psi(x) + \kappa \psi^\dagger(x) \psi^\dagger(x) \psi(x) \psi(x)]
\d{x},
\end{eqnarray*}
where $\kappa>0$ and we have fixed $\hbar=2m=1$.
The bosonic field operators satisfy the canonical commutation relation
$\left[ \psi(x), \psi^\dagger(y) \right] = \delta(x-y)$.
For the $n$-particle state 
$\ket{\Psi_0} = \int \d{x_1}\cdots \d{x_n} \Psi_0(x_1,\ldots,x_n) \psi(x_1)^\dagger \cdots \psi(x_n)^\dagger \ket{0}$, 
one can derive 
the Schr\"odinger equation from the above Hamiltonian: 
\begin{eqnarray}
  \left\{ - \sum_{j=1}^n \partial_{x_j}^2 + 2 \kappa \sum_{1\leq j<j'\leq n} \delta(x_j-x_{j'})\right\}
  \Psi_0
= E\Psi_0
  \label{eq:QNLSsch}
  ,
\end{eqnarray}
where $\Psi_0 = \Psi_0(x_1,\ldots,x_n)$. 
We shall now introduce a lattice version of this model, 
where the spatial position $x\in {\bf R}$ in the continuum model is replaced 
by the site $i\in {\bf Z}$ with the lattice spacing $\Deltax$. 
Let $V_i$ be a physical Hilbert space at the $i$th site spanned by  
$\ket{m}={1\over \sqrt{m!}} (\psi_i^\dagger)^m \ket{0}$ with $m \ge0$. 
Here, $\psi^\dagger_i$ and $\psi_i$ are the bosonic creation and annihilation operators on $V_i$, respectively, 
and they satisfy $\left[\psi_i,\psi_{i'}^\dagger \right] = \delta_{ii'}$, and
$\left[\psi_i,\psi_{i'} \right] = \left[\psi_i^\dagger,\psi_{i'}^\dagger \right]=0$. 
In the continuum limit ($\Deltax \rightarrow 0$), $\psi_i \rightarrow \psi(x) \sqrt{\Deltax}$.
Note that $\psi_i$ and $\kappa \Deltax$ are dimensionless.

Following previous studies\cite{NPB.205.401,TMP.57.1059},
the $L$-operator at the $i$th site is defined by
\begin{eqnarray}
  \calL_{ji}(\lambda) = \left(
    \begin{array}[]{cc}
      1 - {\i \lambda \Deltax \over 2} + {\kappa \Deltax \over 2} \psi_i^\dagger \psi_i
      & -\i \sqrt{\kappa \Deltax} \psi_i^\dagger \rho_i
      \\
      \i \sqrt{\kappa \Deltax} \rho_i \psi_i
      & 1 + {\i \lambda \Deltax \over 2} + {\kappa \Deltax \over 2} \psi_i^\dagger \psi_i
    \end{array}
  \right)
  \nonumber
  ,
\end{eqnarray}
where 
$  \rho_i =\left( 1 + {\kappa \Deltax \over 4} \psi_i^\dagger \psi_i \right)^{1/2}$. 
The matrix elements of $\calL_{ji}$ and $\rho_i$ are operators on $V_i$.
Let us denote the two-dimensional auxiliary space by $\AS_j$,
which is spanned by the two orthonormal states $\ket{\rightarrow}$ and $\ket{\leftarrow}$. 
$\calL_{ji}$ is represented as a $2\times 2$ matrix in $\AS_j$.
This $L$-operator is the same as the classical Lax operator on the lattice 
and satisfies the Yang-Baxter relation: 
$R(\lambda,\lambda') \calL_{ji}(\lambda)
\otimes \calL_{j'i}(\lambda')
= \calL_{j'i}(\lambda')
\otimes \calL_{ji}(\lambda) R(\lambda,\lambda') $
with the $R$-matrix of the Lieb-Liniger model\cite{NPB.205.401}.
The existence of the $R$-matrix ensures the complete integrability of this model 
through the quantum inverse scattering method.
The monodromy matrix is constructed as the following ordered matrix product:
\begin{eqnarray*}
  \calT(\lambda_j) = \prod_{i=1}^N \calL_{ji}(\lambda_j)
  =\left(
    \begin{array}[]{cc}
      \Aa(\lambda_j) & \Ba(\lambda_j)
      \\
      \Ca(\lambda_j) & \Da(\lambda_j)
    \end{array}
  \right)
  ,
\end{eqnarray*}
where
$\calT$ acts on $\AS_j \otimes \mathscr{H}$,
$\mathscr{H}=\otimes_{i=1}^N V_i$ is the total Hilbert space,
and
$N$ is the total number of sites.
The Hamiltonian defined as  ${1\over \i}{\d{}\over \d{\lambda}} \log \tr \calT(\lambda)$ 
is known to be nonlocal (quasi-local)\cite{TMP.57.1059}.
An $n$-particle eigenstate, the Bethe state, is constructed as
\begin{eqnarray}
  \ket{\Psi_a} = \prod_{j=1}^n \Ba(\lambda_{n+1-j})  \ket{0},
  \label{eq:Psia}
\end{eqnarray}
where $\ket{0}$ is the vacuum of the total Hilbert space $\mathscr{H}$, i.e., $\psi_{i} \ket{0} = 0$ for any $i$.
Here, the set of variables $\lambda_j (j=1,\ldots,n)$ corresponding to the
momenta is the solution of the Bethe equation.
Since the Bethe state is our starting point in this letter,
let us skip the Bethe equation and
the algebraic relations among $\Aa, \Ba, \Ca,$ and $\Da$ in the algebraic Bethe ansatz. 

An MPS-type representation of $\ket{\Psi_a}$ itself 
is easily obtained by changing the order of the product of $\calL_{ji}$.
In \refeq{eq:Psia},
first we construct $\Ba$ as
$\Ba(\lambda_j)=\bra{\leftarrow}
\prod_{i=1}^N \calL_{ji}(\lambda_j) \ket{\rightarrow}$,
and then calculate the product of $\Ba(\lambda_j)$.
Conversely,
an MPS representation
is obtained by calculating the product over $j$ before the product over $i$.
Following the notation in ref.~\citen{AX.0911.4215},
the MPS representation is 
given by
\begin{eqnarray}
  \ket{\Psi_a} 
  = {\rm Tr}_{\bar{\mathscr{H}}} 
  \left[
    Q_n \prod_{i=1}^N \left( \calL_{i}(\lambda_{1},\lambda_2,\ldots,\lambda_{n}) \ket{0}\right)
  \right]
  \label{def:Psia:L}
,
\end{eqnarray}
where
$\calL_{i}(\lambda_{1},\lambda_2,\ldots,\lambda_{n}) = \calL_{ni}(\lambda_n) \otimes \cdots \otimes
\calL_{1i}(\lambda_1)$,
$\bar{\mathscr{H}} = \AS_n \otimes \cdots \otimes \AS_1$,
and 
$Q_n = \ket{\rightarrow,
\ldots,\rightarrow}
\bra{\leftarrow,
\ldots,\leftarrow}$
$
=\ket{\Rightarrow}\bra{\Leftarrow}$.
Generally, the MPS is defined by the product of matrices depending on local states.
For $\ket{\Psi_a}$, 
the local matrix for an $m$-particle state at the $i$th site can be defined 
as
\begin{eqnarray}
  C_{n,m}(\lambda_1,\lambda_2,\ldots,\lambda_n) = \bra{m}  \calL_{i}(\lambda_{1},\lambda_2,\ldots,\lambda_{n})
  \ket{0}
  .
\end{eqnarray}
The matrices $C_{n,m},$ and $Q_n$ are $2^n\times 2^n$ matrices acting on $\bar{\mathscr{H}}$.
The recursion relation between $C_{n+1,m}$ and $C_{n,m'}$, 
which becomes important for the following discussion,
is easily obtained as
\begin{eqnarray}
  C_{n+1,m} 
  = \left(
    \begin{array}[]{cc}
      \newb{n+1}{m} C_{n,m} & \newc{m-1} C_{n,m-1} \\
      \newc{m}^*  C_{n,m+1} & \newb{n+1}{m}^* C_{n,m}
      \\
    \end{array}\right)
  \label{eq:recursion}
  ,
\end{eqnarray}
with $C_{n,m}=0 $ ($m<0$ or $m>n$)
and
$C_{0,0}=1$,
where
\begin{eqnarray}
    \newb{j}{m} &=& 1- {\i \lambda_j \Deltax \over 2}  + {\kappa \Deltax \over 2} m
    ,
    \\
  \newc{m} &=& -\i \sqrt{(m+1) \kappa \Deltax\left(1+{m\kappa \Deltax \over 4} \right)}
.
\end{eqnarray}
For the configuration with particles at
$i_1<i_2<\cdots<i_n$,
the component of $\ket{\Psi_a}$
is  given by
\begin{math}
  \Psi_a(x_1,\ldots,x_n) = 
  \Tr_{ \bar{\mathscr{H}}} 
  \left[
    Q_n D_n^{i_1-1} C_n D_n^{i_2 - i_1 -1} C_n\cdots  D_n^{i_n - i_{n-1} -1} C_n D_n^{N-i_n}
  \right]
  ,
\end{math}
where $x_l = i_l a$,
$D_n = C_{n,0}$, and
$C_n = C_{n,1}$.

Let us comment on a similar concept, which is the Suzuki-Trotter (ST) 
decomposition\cite{PRB.31.2957,PTP.56.1454}
in the context of a mapping from a $d$-dimensional quantum system 
into a $(d+1)$-dimensional classical system.
The partition function of the two-dimensional classical system
mapped from the one-dimensional quantum system with $N$ sites
is expressed as 
$Z=\Tr \e^{-\beta H} =\lim_{n\rightarrow \infty} Z_n$
after using the Trotter formula,
where $n$ is the Trotter number
and
$Z_n= \Tr (T_{\rm R})^n = \Tr (T_{\rm V})^N$
with the real-space transfer matrix $T_{\rm R}$
and the virtual-space (quantum) transfer matrix $T_{\rm V}$\cite{PRB.31.2957}.
$T_{\rm V}$ paves the way for analytical and numerical studies:
the thermal Bethe ansatz\cite{PTP.78.1213}
and the finite-$T$ DMRG\cite{PRB.56.5061,JPSJ.66.2221}. 

The similarity becomes clear if we consider a two-dimensional 
statistical model in which the Boltzmann weights are given by $\calL_{ji}$.
For 
the Heisenberg chain, the corresponding model
is a six-vertex model with domain wall boundary conditions (DWBCs)\cite{CMP.86.391}. 
On the other hand, in the present lattice model, 
the partition function 
is defined as the coefficient of the fully filled Bethe state as follows:
$Z_{{\rm DWBC}} = \Psi_a(x_1,x_2,\ldots,x_{N})$. 
Now one can find a clear similarity between 
the MPS of the Bethe state and the ST decomposition,
as shown in Table~\ref{tab:ST}.
The derivation of \refeq{def:Psia:L} from \refeq{eq:Psia} 
also corresponds to that in ST decomposition.
The difference is that $Z_n$ originates from the quantum Boltzmann weight $\e^{-\beta H}$,
while $Z_{{\rm DWBC}}$ originates from the Bethe state $\ket{\Psi_a}$.

Hereafter, we shall derive a continuous MPS
from $\Psi_a$ defined in the discretized real space with the  artificial lattice constant $a$.
This reminds us of
a remarkable advance in the Monte Carlo method, namely,
the continuous (imaginary)-time loop algorithm,\cite{PRL.77.5130}
which completely eliminates the systematic error
due to the artificial discretization with the Trotter number.

\begin{table}
  \centering
  \caption{Correspondence between MPS representation 
of the Bethe state and ST decomposition for an $N$-site system.}
\begin{tabular}{c|c}
  MPS for ABA & ST decomposition
\\\hline
$Z_{{\rm DWBC}}= \Psi_a(x_1,x_2,\ldots,x_{N})$ & $Z_n$
\\
particle number $n$  & Trotter number $n$
\\
$\Ba(\lambda_{j})$ in \refeq{eq:Psia}& $T_{\rm R}$
in $Z_n= \Tr (T_{\rm R})^n $
\\
$\calL_{i}(\lambda_{1},\lambda_2,\ldots,\lambda_{n})$
in \refeq{def:Psia:L} & $T_{\rm V}$
in $Z_n= \Tr (T_{\rm V})^N$ 
\end{tabular}
  \label{tab:ST}
\end{table}

In the continuum limit $\Deltax\rightarrow 0$ with finite $n$,
which corresponds to the situation of \refeq{eq:QNLSsch},
we need to calculate an infinite power of $D_n=C_{n,0}$ to obtain $D_n^{i - i'}=D_n^{ (x - x')/a}$.
Therefore, the diagonalization of $D_n$ is indispensable.
To diagonalize $D_{n}$,
we introduce the invertible $F$-matrices $F_n$
and change the basis in $\bar{\mathscr{H}}$ as
$\Cf_{n,m} = F_n^{-1} C_{n,m} F_n$ and $\Qf_n = F_n^{-1} Q_{n} F_n$.
The state $\ket{\Psi_a}$ is invariant under this global ``gauge'' transformation.
Since $D_{1}$ is diagonal,
we can take the initial $F$-matrix as $F_1=\sigma^0$,
where $\sigma^0$ is the $2\times 2$ identity matrix.
After some calculation and using a general procedure 
(see ref.~\citen{AX.0911.4215}),  
we obtain the following recursive definition for the $F$-matrices:
\begin{eqnarray}
  F_{n+1} = \left(
    \begin{array}{cc}
      F_n & 0\\
      F_n \Ff_n & F_n
    \end{array}
  \right)
  ,
\end{eqnarray}
where
\begin{math}
\Ff_n = \sum_i
{- \i \kappa \over \lambda_i - \lambda_{n+1}}
\Ff_n^{(i)}
\end{math}
and 
$
\Ff_n^{(i)} = \sigma^+_i 
$$
  \prod_{l=i+1}^{n} 
  \left(I_n -{\i \kappa \over \lambda_i - \lambda_{l}} 
  \sigma^z_{l}\right),
$
with
the $2^n \times 2^n$ identity matrix $I_n$
and $2^n \times 2^n$ matrices 
$\sigma^\alpha_{j}=\bigotimes_{l=1}^{n-j} \sigma^0
\otimes \sigma^\alpha
\bigotimes_{l=n-j+2}^{n} \sigma^0$
defined 
by the $2\times 2$ Pauli matrices $\sigma^\alpha$.
It is highly nontrivial from \refeq{eq:recursion}
that
the recursively defined $F_n$ diagonalizes 
a non-Hermitian matrix $C_{n,0}$
and transforms $C_{n,m>0}$ into upper triangular matrices.

Then, we now write down the explicit expression 
\begin{eqnarray}
  \Cf_{n,0} = \Df_n = 
\bigotimes_{l=1}^n \left(
    \begin{array}{cc}
      \newb{n-l+1}{0} & 0 \\
      0 & \newb{n-l+1}{0}^*
    \end{array}\right),
\end{eqnarray}
which is diagonal.
Other matrices are summarized as
\begin{math}
  \Qf_n = Q_n 
\end{math} 
and 
\begin{math}
  \Cf_{n,1} = \sum_{i=1}^{n} \Cf_{n}^{(i)},
\end{math} 
with 
\begin{math}
  \Cf_{n}^{(i)} = {\newc{0} \over \newb{i}{0}} \Cf_{n,0} \Ff_n^{(i)}.
\end{math}
These operators satisfy the following algebraic relations:
\begin{eqnarray}
  \Cf_n^{(j)} \Df_n = z_j \Df_n \Cf_n^{(j)},\;
  \Cf_n^{(j)} \Cf_n^{(k)} = \tilde{S}_{jk}(\lambda_j,\lambda_{k}) \Cf_n^{(k)} \Cf_n^{(j)} 
  \label{eq:CRrel:s-matrix}
,
\end{eqnarray}
where
$z_j = {\newb{j}{0}^* \over \newb{j}{0}}$
and $ 
\tilde{S}_{jj'} = 
  {z_i \over z_j} S_{jj'}
$
with
$S_{jj'} = { \lambda_j - \lambda_{j'} -\i  \kappa
    \over
    \lambda_j - \lambda_{j'}  +\i \kappa }
$.
Moreover,
one can show 
$\Cf_{n,m}  = 
{1\over m!}
\Cf_{n,0} \left( \Cf_{n,0}^{-1} \Cf_{n,1} \right)^m
\prod_{j=1}^{m-1} {\newc{j} \over \newc{0}}
$.

Before we move on to cMPS,
let us comment on the simplification due to the $F$-matrices,
which enables us to make
the five-vertex model
from 
the six-vertex model for the Heisenberg chain\cite{AX.0911.4215}.
For the present model the graphical representation of $Z_{\rm DWBC}$
has infinite nonzero vertices 
because the local Hilbert space $V_i$ has infinite degrees of freedom.
Nevertheless,
by using the $F$-matrices,
we can obtain the five-vertex model under the DWBC.
In this sense,
the present model receives more benefits from the $F$-matrices
than the Heisenberg chain.

Let us consider a diagonal matrix $\Df_{n}^{i_2 - i_1}$ in the continuum limit $\Deltax \rightarrow 0$
with $x_2 - x_1 = (i_2 - i_1) \Deltax$.
Using 
$\lim_{\Deltax\rightarrow 0} \left(1- \i \Deltax {\lambda_j \over 2} \right)^{{1\over \Deltax}}
=\e^{- \i \lambda_j/2}
$, we find
$
  \lim_{\Deltax\rightarrow 0} \Df_{n}^{i_2 - i_1} =   
  \exp\left[
    \i (x_2-x_1) \Kf_n
  \right]$, where 
$\Kf_n =  
-\sum_{j=1}^n {\lambda_j \over 2} \sigma^z_j
$.
Here,
$\e^{\i x \Kf_n}$ can be interpreted as a free propagator.

On the other hand, for $\Cf_{n,m}$ ($m>0$)
one can show the relation 
$\sqrt{m!} \Cf_{n,m} \simeq (\Cf_{n,1})^m$ 
in the continuum limit with a finite $n$.
This guarantees  the continuity of the many-body function, i.e., 
$\Psi(x,x)=\lim_{y\rightarrow x} \Psi(x,y)$.
Changing the basis by using $F$-matrices,
we can show that
$\calL_{i}(\lambda_{1},\lambda_2,\ldots,\lambda_{n})\ket{0}$
becomes
\begin{eqnarray}
  \sum_m \Cf_{n,m} \ket{m} 
  \simeq
  \Df_{n} \exp\left[\Rf_n^\dagger  \psi^\dagger(x) a\right] \ket{0}
  ,
\end{eqnarray}
where $\Rf_n^\dagger=\lim_{a\rightarrow 0} \Df_{n}^{-1}\Cf_{n,1}/\sqrt{a}
=-\i\sqrt{\kappa} \sum_{i=1}^n \Ff_n^{(i)}$ acting on $\bar{\mathscr{H}}$ 
is an analogue of the reflection operator in $\mathscr{H}$\cite{PRL.58.1395}. 

By using the path-ordered operator $P$,
we can write down $\ket{\Psi_0}=\lim_{\Deltax \rightarrow 0} \ket{\Psi_a}$ as
\begin{eqnarray}
  \ket{\Psi_0}
  =
  {\rm Tr}_{ \bar{\mathscr{H}}} 
  \left[
    \Qf_n P  \e^{
        \int  [\i \Kf_n + \Rf_n^\dagger \psi^\dagger(x)] \d{x}
      }
  \right]
  \ket{0}
  .
\end{eqnarray}
The form is seemingly a coherent state with fluctuating particle number,
but $\Qf_n$ projects onto the state with a fixed number of particles. 

\begin{figure}
  \centering
  \resizebox{8cm}{!}{\includegraphics{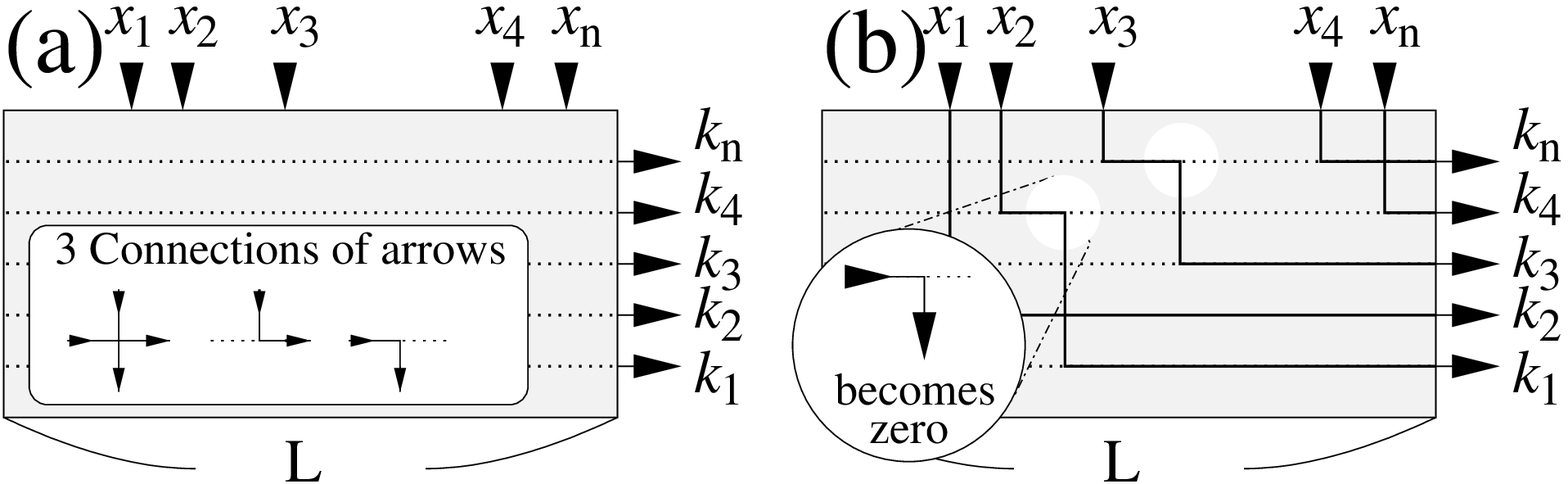}}\\
  \vspace{0.5cm}
  \resizebox{8cm}{!}{\includegraphics{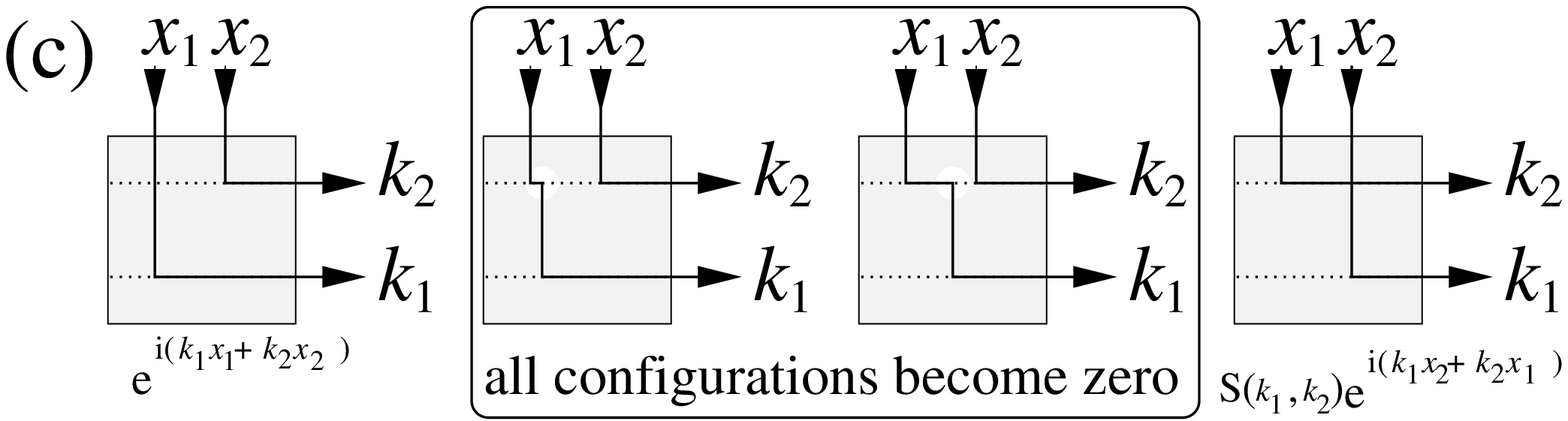}}
  \caption{(a) $n$-particle cMPS,
which is the sum of all configurations
of ``world lines'' between momentum space $k_j=\lambda_j$ and real space $x_i$.
Inset: three connections of world lines.
(b) Configuration of cMPS with vertices that become zero after using $F$-matrices.
(c) Configurations and corresponding weights after using $F$-matrices for a two-particle state.
}
  \label{fig:braid}
\end{figure}

Let us now introduce the graphical representation of the cMPS.
Following the manner of the Heisenberg chain, 
we write the basis of $V_i$ as $\ket{\uparrow}=\ket{0}$ and $\ket{\downarrow}=\ket{1}$.
The $m$-particle state can be disregarded owing to the continuity of the many-body function. 
Unlike in the lattice models, we must draw an infinite number of up arrows for the vacuum in the continuous space. For simplicity, however,
we shall draw only down arrows and right arrows,
as shown in Fig.~\ref{fig:braid}(a).

By using the three kinds of connections  shown in the inset of Fig.~\ref{fig:braid}(a),
the down arrows and right arrows are connected one by one, as shown in 
Fig.~\ref{fig:braid}(b).
We call them ``world lines'' in the continuous space and discrete momentum space,
which is analogous to the world lines in the continuous time and discrete lattice.~\cite{PRL.77.5130}
To calculate $\Psi(x_1,\ldots,x_n)$,
we sum all configurations of world lines.
The number of configurations is infinite, because
we can perform a continuous modification of the configurations,
as shown in Fig.~\ref{fig:braid}(c).
However, 
after using $F$-matrices,
one of the three connections, which corresponds to ``annihilation'',
becomes zero.
Then, we have only two configurations for the two-particle cMPS,
as shown in Fig.~\ref{fig:braid}(c),
that is,
$\Psi(x_1,x_2) \propto
\e^{\i k_1 x_1}
\e^{\i k_2 x_2}
+
S(k_1,k_2)
\e^{\i k_1 x_2}
\e^{\i k_2 x_1}
$.
The simplification due to the $F$-matrices
is valuable for developing a new numerical method.

In the graphical representation,
the horizontal line corresponds to 
the plane-wave-type function $\e^{\i k x}$.
It originates from the free propagator $\e^{\i \Kf_n x}$ except for the overall factor,
where $\Kf_n$ is interpreted as a momentum operator.
The meaning of the scattering matrix $S(k_1,k_2)=S_{12}(k_1,k_2)$
is clarified by crossed world-lines in Fig.~\ref{fig:braid}(c).
It originates from the algebraic relation \refeq{eq:CRrel:s-matrix} and $\lim_{a\rightarrow 0} \tilde{S}_{jj'} = S_{jj'}$.
In short, 
the permutation of plane waves gives the scattering matrix
as in the coordinate Bethe ansatz.
Algebraic relations among ${\cal Q}_n, \Kf_n,$ and $\Ff_n^{(i)}$, which define $\Rf_n$,
are important in the continuous matrix product ansatz
and
can be obtained generally
as a natural extension
of those in the matrix product ansatz\cite{Alcaraz}.
For an extension of the statistical model,
a corner transfer matrix (CTM) of the cMPS is obtained as a mapping from momentum space into real space.
Since the original CTM\cite{\CTM} is interpreted as a Lorentz boost
\cite{PD.18.348,PTP.69.431},
it is an interesting problem to study the field theory and algebraic structure behind 
the CTM of cMPS.

Finally, we comment on the cMPS defined by Verstraete and Cirac\cite{AX.1002.1824} as a variational state:
$\ket{\Phi}=\Tr
  \left[
    P \left( \exp\left[
        \int  M_0(x) + M_1(x) \psi^\dagger(x) \d{x}
      \right]
    \right)
  \right]
  \ket{0},$
where $M_0(x)$ and $M_1(x)$ are variational matrices with finite dimension $d$
under the assumption of translational invariance for $M_0$ and $M_1$, respectively.
The exact cMPS in the present letter 
shows finite dimensionality $d=2^n$ and 
the translational invariance $M_0(x)=\i \Kf_n$ and $M_1(x)=\Rf_n$.
A significant difference is that while $\ket{\Phi}$ shows a particle number fluctuation, the exact cMPS $\ket{\Psi}$ 
has a fixed number of particles owing to the existence of $\Qf_n$.
It is historically interesting that
the generalized variational MPS\cite{PRL.75.3537}
has the boundary operator $\Qf$.
In short, the exact cMPS has the operator $\Qf_n$ fixing the total number of particles
as the ``boundary'' condition in the auxiliary space $\bar{\mathscr{H}}$.
To tackle future problems such as a finite-temperature or higher-dimensional generalization,
the physical meaning of the matrices revealed in the letter would become more important.

\acknowledgement

The authors are grateful to M. Suzuki for his valuable comments and discussion. 
This work was supported in part by a Grant-in-Aid (No. 20740214) from the Ministry of Education, Culture, Sports, Science and Technology of Japan. 
HK is supported by the JSPS Postdoctoral Fellowships for Research Abroad.

\end{document}